\documentclass[10pt]{article}
\usepackage[english]{babel}
\usepackage{makeidx}
\usepackage{epsfig}

\usepackage{graphicx}
\usepackage{dcolumn}
\usepackage{amsmath}
\usepackage{amssymb}

\usepackage{overcite}
\def\erfc{{\rm erfc}}
\def\erf{{\rm erf}}
\def\rv{{\bf r}}
\def\sv{{\bf s}}

\def\Rv{{\bf R}}

\def\beq{\begin{equation}}
\def\eeq{\end{equation}}

\def\ft{f}

\def\vt{\tilde{v}}

\begin{document}
\title{The high-density limit of two-electron systems: \\ Results
from the extended Overhauser approach}
 \author{Paola Gori-Giorgi and Andreas Savin \\
{\it Laboratoire de Chimie Th\'eorique,} \\
{\it CNRS UMR7616, Universit\'e Pierre et Marie Curie, } \\
{\it 4 Place Jussieu,
F-75252 Paris, France}}
\date{\today}

\maketitle

\begin{abstract}
The ``extended Overhauser model'' [Overhauser, A.W.
{\it Can. J. Phys.} {\bf  1995}, {\it 73}, 683] for the calculation
of the spherically and system-averaged pair density (APD) has been 
recently combined with the  Kohn-Sham equations to 
yield realistic APD and correlation energies.
In this work we test this approach in the
high-density (weakly-correlated) limit of the He isoelectronic series
and of the Hooke's atom isoelectronic series.
Unlike many of the commonly used energy functionals, the
Overhauser approach
yields accurate correlation energies for both series. 
\end{abstract}

\section{Introduction}
Kohn-Sham (KS) Density Functional Theory\cite{kohnnobel,science,FNM}
(DFT) is nowadays one of the most
popular methods for electronic structure calculations both in chemistry
and solid-state physics, thanks to its combination
of low computational cost and reasonable performances.
The accuracy of a KS-DFT result is limited
by the approximate nature of the exchange-correlation energy
density functional $E_{xc}[n]$. Simple approximations 
(local-density approximation and generalized gradient corrections)
for $E_{xc}[n]$ provide 
practical estimates of thermodynamical, structural and spectroscopic properties
of atoms, molecules and solids. However, with the current approximations,
KS-DFT is still lacking in several 
aspects, in particular it fails to handle near-degeneracy correlation 
effects (rearrangement of electrons within partially filled shells) and to 
recover long-range van der Waals interaction energies. 
The inaccuracy of KS-DFT stems from our lack of knowledge of $E_{xc}[n]$,
 and much effort is put nowadays in finding new
approximations to this term (for recent reviews, see, e.g.,
Refs.~\citen{science,FNM,prescription}). A popular
trend in the development of new KS $E_{xc}[n]$
is the use of
the exact exchange functional $E_x[n]$ (in terms of the KS orbitals), and
thus the
search for an approximate, compatible, correlation functional $E_c[n]$.

An exact expression for $E_c[n]$ is the coupling-constant 
integral\cite{gunnarsson,wang}
\beq
E_c[n]=\int_0^{\lambda_{\rm phys}} d\lambda \int_0^\infty dr_{12}\, 
4\pi\,r_{12}^2\,f_c^{\lambda}(r_{12})\frac{\partial w^{\lambda}(r_{12})}
{\partial \lambda},
\label{eq_Ecfromf}
\eeq 
where the interaction between the electrons is adiabatically turned on
from $w^{\lambda=0}(r_{12})=0$ to the Coulomb repulsion 
$w^{\lambda=\lambda_{\rm phys}}(r_{12})=1/r_{12}$ by varying a real
parameter $\lambda$ (typical examples are $w^\lambda(r_{12})=\lambda/r_{12}$,
with $\lambda_{\rm phys}=1$,  or 
$w^\lambda(r_{12})=\erf(\lambda r_{12})/r_{12}$,
with $\lambda_{\rm phys}=\infty$). The one-electron density $n(\rv)$ is
(ideally) kept independent of $\lambda$ and
equal to the one of the physical system
by means of a suitable external potential $v^\lambda(\rv)$. In 
Eq.~(\ref{eq_Ecfromf}) the correlation part of the spherically and 
system-averaged pair density (intracule density) 
$f_c^\lambda(r_{12})$ is defined as follows.
For each $\lambda$, take the square of the 
many-electron wavefunction $\Psi^\lambda$
ground-state of the hamiltonian $H^\lambda$,
\beq
H^\lambda=-\sum_{i=1}^N\frac{\nabla_i^2}{2}+\frac{1}{2}\sum_{i\neq j=1}^N
w^\lambda(|\rv_i-\rv_j|)+\sum_{i=1}^N v^\lambda(\rv_i),
\eeq
and integrate it
over all variables but the scalar electron-electron distance
$r_{12}=|\rv_1-\rv_2|$,
\beq
f^\lambda(r_{12}) = \frac{N(N-1)}{2}\sum_{\sigma_1...\sigma_N}
 \int |\Psi^\lambda(\rv_{12},\Rv,\rv_3,...,\rv_N)|^2
\frac{d\Omega_{\rv_{12}}}{4\pi} d\Rv d\rv_3...d\rv_N,
\label{eq_intra}
\eeq
where $\Rv=(\rv_1+\rv_2)/2$. The correlation part $f_c^\lambda(r_{12})$
is then defined as $f_c^\lambda(r_{12})=f^\lambda(r_{12})-f_{\rm KS}(r_{12})$,
where the intracule density of the KS system is 
$f_{\rm KS}(r_{12})=f^{\lambda=0}(r_{12})$ (and yields the Hartree
plus the exchange energy).

The traditional DFT approach to the construction of approximate $E_c[n]$
is based on the idea of universality. 
For example, the familiar local-density
approximation (LDA) consists in transfering,
in each point of space,
the pair density from the uniform electron gas to obtain
an approximation for $f_c^\lambda(r_{12})$ in Eq.~(\ref{eq_Ecfromf}). 
In a couple of recent papers, \cite{GS1,GS2,GS4}
we have started to explore a different way of constructing
$E_c[n]$, based on an ``average pair density functional theory''
(APDFT), which was inspired by the seminal work of Overhauser \cite{Ov}
and its subsequent extensions. \cite{GP1,pisani,CGPenagy2} 
In this approach, we solve a set of
radial (one-dimensional) Schr\"odinger-like equations
that give, in principle, the exact $f^\lambda(r_{12})$ along the DFT
adiabatic connection. In practice, this formalism contains an
unknown effective electron-electron interaction that needs to be 
approximated. The APDFT 
equations must be solved for each system,
and combined self-consistently with the KS equations.\cite{GS4}
Preliminary applications of this approach, combined with a
 simple approximation \cite{GS1}
 for the effective electron-electron interaction
that enters in the formalism, gave accurate intracule densities
$f(r_{12})$ and correlation energies $E_c[n]$ for the He isoelectronic
series. \cite{GS1,GS4}

Katriel {\it et al.} \cite{katriel} have recently tested
most of the currently available correlation energy functionals
in the high-density (weakly-correlated)  limit of the He and
of the Hooke's atom isoelectronic series, finding that, while several
functionals are accurate for the He sequence, none is satisfactory for
the Hooke's atom series. Motivated by their findings, in this
work we compute the correlation energy and the intracule density
in the high-density limit of the two series via the APDFT
approach, \cite{GS1,GS2,GS4} finding accurate
results in both cases.

The paper is organized as follows. In the next Sec.~\ref{sec_wcl} we 
recall the basic equations that define the high-density
limit of the He and Hooke's atom sequences, to which we apply,
in Sec.~\ref{sec_eff} and  \ref{sec_Ec},
the formalism of Refs.~\citen{GS1,GS2,GS4}  to
compute the intracule density and the correlation energy.
In Sec.~\ref{sec_LDA} we also analyze the failure of LDA  in this limit
from the point of view of $f(r_{12})$. The last Sec.~\ref{sec_conc} is
devoted to conclusions.

\section{The high-density limit of the He and Hooke's atom isoelectronic
series}
\label{sec_wcl}
The two hamiltonians analyzed in this paper read
\begin{eqnarray}
H & = & -\frac{\nabla^2_1}{2}-\frac{\nabla^2_2}{2}+v(r_1)+v(r_2)+\frac{1}{r_{12}}, \\
v(r) &  = & \left\{
\begin{array}{ll} 
-\frac{Z}{r} & \qquad \qquad {\rm He\; series}  \\
\frac{1}{2}\,k \,r^2 & \qquad \qquad {\rm Hooke's\; atom\; series.} 
\end{array}
\right.
\nonumber
\end{eqnarray}
We are interested in the high-density (weakly-correlated) limit,
which corresponds to $Z\to\infty$ and $k\to\infty$. By switching to
scaled coordinates $\sv=\rv/\alpha$, with  $\alpha=Z^{-1}$ (He series) and
$\alpha=k^{-1/4}$ (Hooke's series), both hamiltonians have the form
\beq
H  =  \frac{1}{\alpha^2}\left(-\frac{\nabla^2_{\sv_1}}{2}-\frac{\nabla^2_{\sv_2}}{2}+\vt(s_1)+\vt(s_2)+\frac{\alpha}{s_{12}}\right)\equiv \frac{1}{\alpha^2} (\tilde{H}_0+\alpha \tilde{H}_1), 
\label{eq_Hgeneral}
\eeq
where $\vt(s)=-1/s$ for the He series, and 
$\vt(s)=s^2/2$ for the Hooke's atom series. We 
thus study pertubatively
the system described by $\tilde{H}_0+\alpha \tilde{H}_1$.

The order zero of the one-electron density $n(r)$ and of
the intracule density $f(r_{12})$, in scaled units, is simply
\begin{eqnarray}
n^{(0)}(s) & = & \left\{
\begin{array}{ll}\frac{2}{\pi}e^{-2\,s} & ({\rm He})\\
 \frac{2}{\pi^{3/2}}e^{-s^2} & ({\rm Hooke})
\end{array}
\right. \\
f^{(0)}(s_{12}) & = &  \left\{
\begin{array}{ll}
\frac{1}{24\pi} (3+6\, s_{12}+4\, s_{12}^2)\, 
e^{-2\,s_{12}} & ({\rm He}) \\
 \frac{1}{(2\pi)^{3/2}} e^{-s^2_{12}/2} & ({\rm Hooke})
\end{array}
\right. 
\end{eqnarray}
These functions are correctly normalized, so that if we switch back
to coordinates $\rv$ we have $n^{(0)}(r)=\alpha^{-3} \,n^{(0)}(s=\alpha^{-1}r)$, etc.
\par
The first-order correction to the scaled density, $n(s)=
n^{(0)}(s)+
\alpha\, n^{(1)}(s)+...$ is given by
\beq
n^{(1)}(s) = 2 n^{(0)}(s) \chi(s),
\label{eq_densord1}
\eeq
where~\cite{schwartz}
\beq
\chi(s)  =  -\frac{23}{32}-\frac{e^{-2 s}}{4}-\frac{3}{8}\gamma
+\frac{3}{16}\frac{1-e^{-2s}}{s}+\frac{5}{8} s+\frac{3}{8}
{\rm Ei}(-2s)-\frac{3}{8}\ln(s),
\label{eq_chi}
\eeq
for the He isoelectronic series, with $\gamma=0.577216..$,
\beq
{\rm Ei}(-x)=-\int_x^{\infty} \frac{e^{-t}}{t} dt,
\eeq
and \cite{kasia}
\begin{eqnarray}
\chi(s)& = & \frac{\erf(s)}{s}-\frac{\sqrt{2}(1+\ln 2)}{\sqrt{\pi}}-
\frac{1}{s\sqrt{\pi}}\int_0^\infty dx\,
\left(e^{-(x-s)^2}-e^{-(x+s)^2}\right)
\nonumber \\ 
& \times & \left[e^{x^2/2}\erfc\left(
\frac{x}{\sqrt{2}}\right) 
 +\sqrt{2}\, x\int_0^{x/\sqrt{2}} dt\,e^{t^2}
\erfc(t)\right],
\end{eqnarray}
for the Hooke's atom isoelectronic series.

By definition, the Kohn-Sham hamiltonian describes a non-interacting
system that has the same density of the physical, interacting, system. 
Thus, the 
first-order change in the electron density of Eq.~(\ref{eq_densord1})
corresponds to a first-order change in the KS system.
Therefore, we write the  scaled intracule $f(s_{12})$ 
up to orders $\alpha$ as
\beq
f(s_{12})=f^{(0)}(s_{12})+\alpha 
\left[f_{\rm KS}^{(1)}(s_{12})+
f_c^{(1)}(s_{12})\right]+O\left(\alpha^2\right),
\label{eq_expf}
\eeq
where we have separated the first-order correction into a 
Kohn-Sham part and a correlation part. The KS part
$f_{\rm KS}^{(1)}$ is entirely determined by the first-order density 
$n^{(1)}$ of Eq.~(\ref{eq_densord1}),
\beq
f_{\rm KS}^{(1)}(s_{12})=\int d\sv\,n^{(0)}(\sv+\sv_{12})\, 
n^{(0)}(s)\,\chi(s),
\label{eq_fKS1}
\eeq
and is reported in Appendix~\ref{app_fKS1} in analytic form
for the He isoelectronic
series, while is obtained numerically for the case of the Hooke's series.

The total first-order intracule $f^{(1)}=f_{\rm KS}^{(1)}+
f_c^{(1)}$ is known
analytically in the case of the Hooke's series,\cite{kasia}
\beq
f^{(1)}(s_{12})  =  \frac{2\,e^{-s^2_{12}/4}}{(2\pi)^{3/4}}
\biggl[1-\frac{1+\ln 2}{\sqrt{2\pi}}+\frac{1}{s_{12}}-\frac{e^{s^2_{12}/2}}{s_{12}}
\erfc\left(\frac{s_{12}}{\sqrt{2}}\right)
 +  \sqrt{2}\int_0^{s_{12}/\sqrt{2}}e^{t^2}\erfc(t)\,dt\biggr].
\eeq

\section{Effective equations for $f(r_{12})$ in the high-density limit}
\label{sec_eff}
\subsection{Formalism}
We are interested in calculating $f_c^{(1)}$ 
and the corresponding second-order correlation energy $E_c^{(2)}$
with the method of Refs.~\citen{GS1,GS2,GS4}, 
in which the intracule density $f(r_{12})$ of the physical system is obtained
from a set of effective equations, which for two-electron systems
reduce to
\beq
[-\nabla_{r_{12}}^2+w_{\rm eff}(r_{12})]\, \psi(r_{12})=\epsilon\, \psi(r_{12}),
\label{eq_eff}
\eeq
with $f(r_{12})=|\psi(r_{12})|^2$. Equation~(\ref{eq_eff}) can be derived
by considering\cite{GS2,GS4} a set of Hamiltonians characterized by a
real parameter $\xi$,
\beq
H^\xi= -\sum_{i=1}^N\frac{\nabla_i^2}{2}+\frac{1}{2}\sum_{i\neq j=1}^N
w^\xi(|\rv_i-\rv_j|)+\xi\sum_{i=1}^N v_{ne}(\rv_i), \qquad
f^\xi(r_{12})=f(r_{12})\;\forall \xi
\label{eq_adiaAPDFT}
\eeq
that describe a set of systems in which 
the external potential is turned off
as $\xi\to 0$, and the intracule density is kept fixed, equal to the one
of the physical system, by means of a suitable electron-electron interaction
$w^\xi(r_{12})$. In the case $N=2$, when $\xi=0$ 
we have a translationally-invariant
system (the center-of-mass degree of freedom is described by a plane wave) 
of two fermions in a relative bound 
state (similar
to the case of positronium, but with a different interaction). This 
relative bound state
is such that the square of the wavefunction for the relative coordinate
$r_{12}$ is equal to $f(r_{12})$ of the starting physical 
system, and is thus described by Eq.~(\ref{eq_eff}).\cite{GS2,GS4} 
For more than two electrons, in the case of a confined system 
(atom, molecule),
the limit $\xi\to 0$ in Eq.~(\ref{eq_adiaAPDFT})
describes a cluster of fermions, and Eq.~(\ref{eq_eff}) becomes an 
approximation\cite{GS2,GS4,nagy} for the internal degrees of freedom
of the cluster. 

Here we focus on the high-density limit of the hamiltonians of
Eq.~(\ref{eq_Hgeneral}) and we thus stick to the case $N=2$. In general,
the effective electron-electron interaction $w_{\rm eff}(r_{12})$ 
in Eq.~(\ref{eq_eff}) is not known, and must be 
approximated. In the case of the He series, we have found\cite{GS1,GS2,GS4}
that a simple approximation based on the original idea of
Overhauser\cite{Ov,GP1} gives very accurate results
for $2\le Z\le 10$. In what follows we analyze the performance of
the same approximation in the very $Z\to\infty$ limit, and we
extend our study to the $k\to\infty$ limit of the Hooke's atom series. 
Of course, in the special case of the Hooke's series, 
the hamiltonian (\ref{eq_Hgeneral}) is exactly separable into
center-of-mass and relative coordinates, so that the exact
$w_{\rm eff}(r_{12})$ is directly available. However, the point here is
to check whether the same approximate $w_{\rm eff}(r_{12})$
that accurately describes the He series is capable to yield also
good results for the Hooke's series, since this seems to be not
the case for the currently available correlation energy 
functionals.\cite{katriel}

The construction of an approximation for the e-e effective 
potential $w_{\rm eff}$ starts with the decomposition \cite{GS1,GS2,GS4}
\beq
w_{\rm eff}(r_{12})=w_{\rm eff}^{\rm KS}(r_{12})+w_{\rm eff}^c(r_{12}),
\label{eq_wdecomp}
\eeq
where $w_{\rm eff}^{\rm KS}=\nabla^2\sqrt{f_{\rm KS}}/\sqrt{f_{\rm KS}}$ 
is the potential that generates the Kohn-Sham
$f_{\rm KS}$ via Eq.~(\ref{eq_eff}), and $w_{\rm eff}^c(r_{12})$
is a correlation potential that needs to be approximated. In the
usual DFT language, Eq.~(\ref{eq_wdecomp}) implies that we are
treating exchange exaclty. 

In scaled units $\sv$, using standard perturbation theory we obtain
the equation for the first-order contribution to $f$
[see Eq.~(\ref{eq_expf})], that separates
into the Kohn-Sham and the correlation parts:
\begin{eqnarray}
\left[-\nabla^2+w_{\rm eff}^{\rm KS\,(0)}-\epsilon^{(0)}\right]
\psi^{(1)}_{\rm KS}  =  \left[\epsilon_{\rm KS}^{(1)}
-w_{\rm eff}^{\rm KS\,(1)}\right] \psi^{(0)}
\label{eq_KS1} \\
\left[-\nabla^2+w_{\rm eff}^{\rm KS\,(0)}-\epsilon^{(0)}\right]
\psi^{(1)}_c  =  \left[\epsilon_{c}^{(1)}
-w_{\rm eff}^{c\,(1)}\right] \psi^{(0)}, 
\label{eq_corr1}
\end{eqnarray} 
where $\psi^{(0)}=\sqrt{f^{(0)}}$, $f_{\rm KS}^{(1)}=
2\psi^{(0)}
\psi^{(1)}_{\rm KS}$, $f_c^{(1)}=2\psi^{(0)}\psi^{(1)}_{c}$,
and
\begin{eqnarray}
w_{\rm eff}^{\rm KS\,(0)}(s_{12}) & = & \frac{2\,(8\,s_{12}^4-
8\,s_{12}^3-38\,s_{12}^2-36\,s_{12}
-9)}{(4\,s_{12}^2+6\,s_{12}+3)^2}-1 \;\; {\rm (He\;series)} \\
w_{\rm eff}^{\rm KS\,(0)}(s_{12}) & = & \frac{s_{12}^2}{4}
\qquad \qquad \qquad \qquad \qquad \qquad \qquad {\rm (Hooke's\;series).}
\end{eqnarray}
In Eq.~(\ref{eq_KS1}), $f_{\rm KS}^{(1)}$ is exatly known
for both series, 
so that we can also obtain $w_{\rm eff}^{\rm KS\,(1)}$ by inversion.
\par
We thus concentrate on the correlation part, since we want to test
approximations for $w_{\rm eff}^{c}$. 
Defining $u_c(x)=x\psi^{(1)}_{c}(x)$ and
$u_0(x)=x\psi^{(0)}(x)$, we have
\beq
\left[\frac{d^2}{dx^2}-w_{\rm eff}^{\rm KS\,(0)}+\epsilon^{(0)}\right]
u_c=\left[w_{\rm eff}^{c\,(1)}-\epsilon_{c}^{(1)}\right]u_0.
\eeq
Following the method of Refs.~\citen{schwartz,dalgarno,march} we look for
a solution of the kind $u_c(x)=u_0(x) \,y(x)$. The function $y(x)$ is then
given by
\beq
y(x)=\int_0^x\frac{dx'}{u_0^2(x')}\int_0^{x'} u_0^2(x'') 
[w_{\rm eff}^{c\,(1)}(x'')-\epsilon_{c}^{(1)}]\,dx''  +C_2.
\label{eq_y}
\eeq
The constant $C_2$ is fixed by requiring the proper normalization,
\beq 
\int_0^{\infty}f_c^{(1)}(x)\, x^2 dx=0 \;\; \Rightarrow \;\;
\int_0^{\infty} y(x) u_0^2(x)\,dx=0.
\eeq
The other integration constant has been fixed in Eq.~(\ref{eq_y}) by
setting equal to zero an unphysical term $C_1 \int^x u_0^{-2}(x') dx'$ that
would make $u_c(x)$ diverge for large $x$.
\begin{figure}
\begin{center}
\includegraphics[width=7.cm]{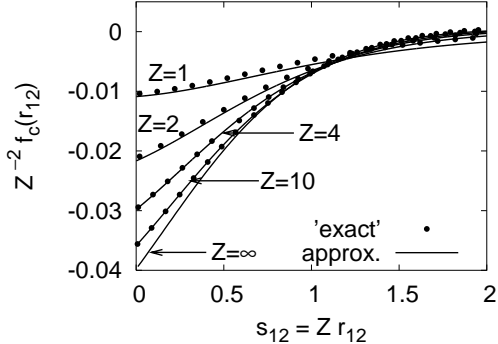}  
\end{center}
\caption{The correlation part of the intracule density, $f_c=f-f_{\rm KS}$,
 divided by $Z^2$, as a function
of the scaled variable $s_{12}=Z r_{12}$ for the He isoelectronic series. 
The ``exact'' results are obtained
from the accurate wavefunctions of Ref.~\citen{morgan}. 
Approximate results
at finite $Z$ using the ``Overhauser  model'' are taken from
Ref.~\citen{GS1}. The $Z=\infty$ result corresponds to 
Eq.~(\ref{eq_y}) with the potential of Eq.~(\ref{eq_ovscaled}).}
\label{fig_fcall}
\end{figure}

\subsection{Testing approximations: the Overhauser potential}
In Refs.~\citen{GS1,GS2,GS4} 
an approximation for $w_{\rm eff}^c$ was built
as an average ``Overhauser-type'' potential,\cite{Ov,GP1}
\beq
w_{\rm eff}^c(r_{12})\approx \left(\frac{1}{r_{12}}+
\frac{r_{12}^2}{2\,\overline{r}^3_s}-\frac{3}{2\,\overline{r}_s}\right)
\theta\left(\overline{r}_s-r_{12}\right),
\label{eq_ov}
\eeq
where $\theta(x)$ is the Heaviside step function and $\overline{r}_s$ is
related to the average density, or, better to the dimension
of the system. For two-electron atoms it was simply estimated as\cite{GS1}
\beq
\overline{r}_s=\left(\tfrac{4 \pi}{3}\,\overline{n}\right)^{-1/3},
\label{eq_avrs}
\eeq
where
\beq
\overline{n}=\frac{1}{N}\int d\rv\,n(\rv)^2.
\label{eq_avn}
\eeq
The idea beyond this approximation is the following. The
e-e correlation potential $w_{\rm eff}^c(r_{12})$
changes the Kohn-Sham
$f$ into the physical one, and must thus
keep the information on the one-electron density (which is the
same in the two systems) while turning on the e-e interaction $1/r_{12}$. In 
Eqs.~(\ref{eq_ov})-(\ref{eq_avn}) this information is 
approximately kept via the average density $\overline{n}$.
\par
In scaled units, the Overhauser potential to first order in $\alpha$,
to be used in Eq.~(\ref{eq_corr1}), becomes
\beq
w_{\rm eff}^{c\,(1)}(s_{12})\approx  \left(\frac{1}{s_{12}}+\frac{s_{12}^2}{2\,\overline{s}^3_s}-\frac{3}{2\,\overline{s}_s}\right)
\theta\left(\overline{s}_s-s_{12}\right),
\label{eq_ovscaled}
\eeq
where, if we adopt the prescription of
Eqs.~(\ref{eq_avrs})-(\ref{eq_avn}), $\overline{s}_s=3^{1/3}+O(\alpha)$ for
the He series and $\overline{s}_s=(3\sqrt{\pi})^{1/3}+O(\alpha)$ for
the Hooke's atom series.

\begin{figure}
\begin{center}
\includegraphics[width=7.cm]{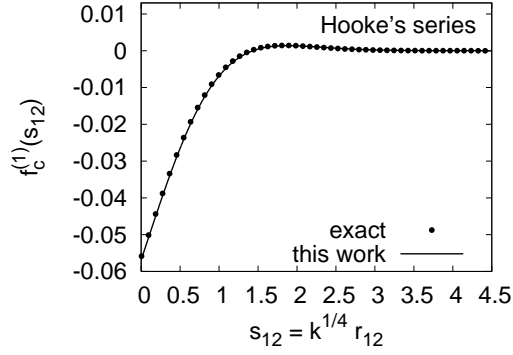}  
\end{center}
\caption{The correlation part of the 
first-order intracule, $f_c(s_{12})$
[see Eq.~(\ref{eq_expf})], for the Hooke's series. The
exact values are compared with the results from the 
Overhauser-type approximation of Eq.~(\ref{eq_ovscaled}).}
\label{fig_fc1Hooke}
\end{figure}

Equation~(\ref{eq_y}) with the potential of Eq.~(\ref{eq_ovscaled}) can 
be evaluated analytically as a function of $s_{12}$ and $\overline{s}_s$
for both series,
although the final expressions are cumbersome and will not be reported here. 
The resulting
$f_c^{(1)}$ for the He series
is shown in Fig.~\ref{fig_fcall}, together
with the corresponding scaled quantity, $Z^{-2} f_c(s/Z)$, for some finite 
$Z$.
[Since $f_c^{(1)}(s)=\lim_{Z\to\infty} Z\,f_c(s)$, and
$f_c(s)=Z^{-3} f_c(s/Z)$, the quantity
to be compared with $f_c^{(1)}(s)$ is $Z^{-2} f_c(s/Z)$.]
For finite $Z$ we show both the ``exact'' 
result~\cite{morgan}
and the approximate result~\cite{GS1} obtained with the Overhauser-type 
potential of Eqs.~(\ref{eq_ov})-(\ref{eq_avn}). 
We see that the $Z$ dependence of the
 short-range part of $f_c$ is very well captured by this simple approximation.
Figure~\ref{fig_fcall} also suggests that the $Z\to\infty$ limit of the
short-range part of $f_c$ is well described by this approach. 
In Fig.~\ref{fig_fc1Hooke} we show the result for $\ft_c^{(1)}$
in the case of the Hooke's series from the Overhauser potential 
compared to the exact one, finding very accurate agreement.

\begin{figure}
\begin{center}
\includegraphics[width=7.cm]{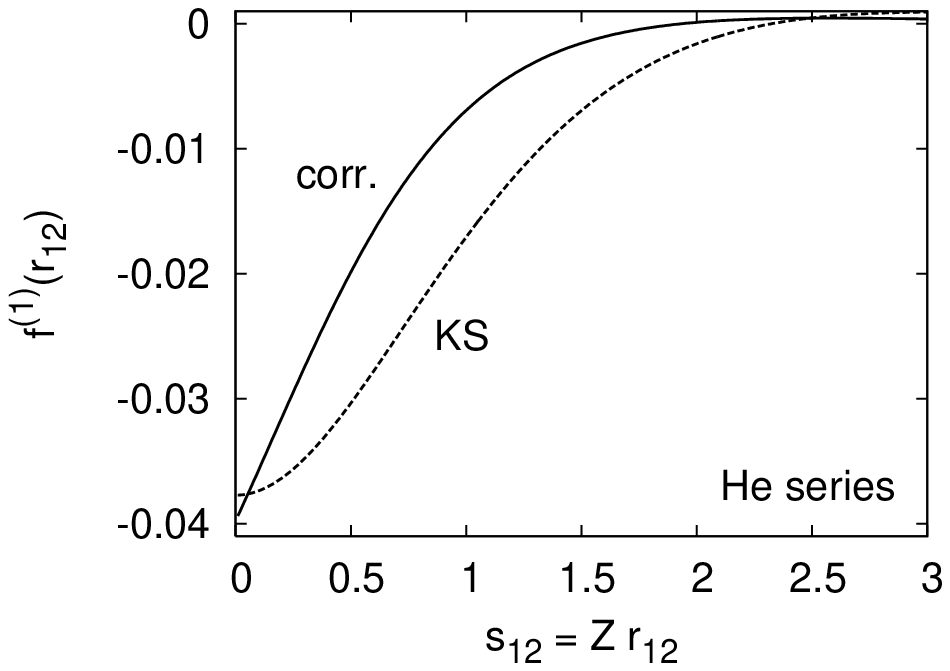}  

\includegraphics[width=7.cm]{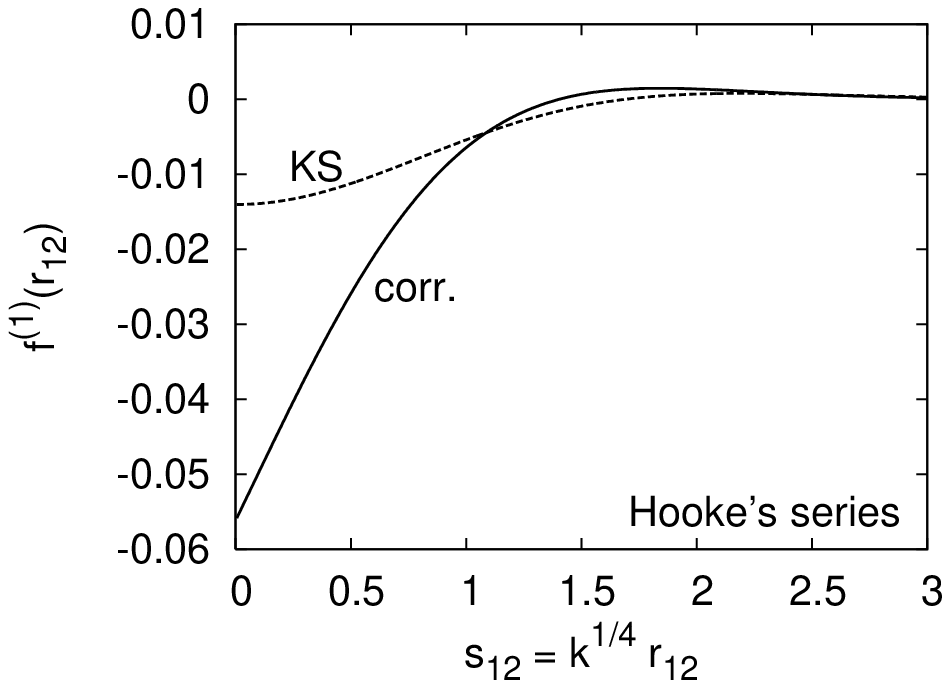}  
\end{center}
\caption{The decomposition of the first-order intracule intracule 
density $\ft^{(1)}(s_{12})$ 
[see Eq.~(\ref{eq_expf})]: the Kohn-Sham part and
the correlation part.}
\label{fig_ord1}
\end{figure}

The KS and the correlation components of $\ft^{(1)}$ are shown
in Fig.~\ref{fig_ord1} for both series. We see that in the case of the
He series the KS and the correlation parts have roughly the
same depth, while in the case of the Hooke's series the 
correlation part is much deeper than the KS one. 
This is due to the fact that the KS part gives the 
change in the e-e distance probability distribution
only due to the first-order change in the one-electron density. 
In the case of the Hooke's series the first-order change in the density
is much smaller, since the harmonic confining external potential 
is stronger than the Coulombic one. Indeed, the function $\chi(s)$
of Eq.~(\ref{eq_densord1}) in the case of the He series is about
twice the one for the Hooke's atom series.

\section{Adiabatic connection and correlation energy}
\label{sec_Ec}
The APD $\ft_c^{(1)}(s_{12})$ gives
the correlation contribution to second order to the expectation 
$\langle V_{ee}\rangle$ of the Coulomb
electron-electron repulsion operator, $V_{ee}=1/r_{12}$, 
\beq
\langle V_{ee}\rangle = \frac{1}{\alpha^2}\left[\alpha \langle 
V_{ee}\rangle ^{(1)}+ \alpha^2 \langle V_{ee}\rangle^{(2)}+O(\alpha^3)\right],
\eeq
where $\langle V_{ee}\rangle ^{(2)} = 
\langle V_{ee}\rangle^{(2)}_{\rm KS}+\langle 
V_{ee}\rangle^{(2)}_c$, and
\beq
\langle V_{ee}\rangle_c^{(2)}
=\int_0^\infty 4\pi\, s_{12}\,\ft_c^{(1)}(s_{12})\, ds_{12}.
\eeq
Our $\ft_c^{(1)}$ from the Overhauser potential give
 $\langle V_{ee}\rangle_c^{(2)}=-0.10256$~Ha for the He sequence
(to be compared with the exact \cite{HU} value,
$ -0.09333$~Ha), and $\langle V_{ee}\rangle_c^{(2)}=
-0.10377$~Ha for the Hooke's series
(to be compared with the exact \cite{kasia} value,
$- 0.09941$~Ha). The error is thus 9~mH for the He
series and 4~mH for the Hooke's series.

The correlation energy can then be otbained via
the adiabatic connection formula of Eq.~(\ref{eq_Ecfromf}), which
for $E_c^{(2)}$ reads
\beq
E_c^{(2)}=\int_0^{\lambda_{\rm phys}} d\lambda \int_0^\infty ds_{12}\, 
4\pi\,s_{12}^2\,\ft_c^{\lambda\;(1)}(s_{12})\frac{\partial w^{\lambda}(s_{12})}
{\partial \lambda},
\label{eq_adiabatic}
\eeq
where $\ft_c^{\lambda\;(1)}$ is the first-order correlated part of 
$\ft$ for the system with interaction 
$\alpha\, w^{\lambda}(s_{12})$. If we were
able to calculate the exact  $\ft_c^{\lambda\;(1)}$ for any $w^{\lambda}$,
the resulting $E_c^{(2)}$ from Eq.~(\ref{eq_adiabatic}) would be independent
of the choice of $w^{\lambda}$. However, when we deal with approximations,
we can obtain better results with some choices rather than others.

As in Ref.~\citen{GS1}, we build an Overhauser-type potential along
the adiabatic connection as
\beq
w_{\rm eff}^{c,\,\lambda}(s_{12};\overline{s}_s)=w^{\lambda}(s_{12})
-
\int_{|\sv|\le \overline{s}_s} 
\overline{n}\,  w^{\lambda}(|\sv - \sv_{12}|)\,d \sv,
\label{eq_vefflambda}
\eeq
where, in scaled units, if we stick with the choice of 
Eqs.~(\ref{eq_avrs})-(\ref{eq_avn}),
$\overline{n}=(4\pi)^{-1}$ for the He series and
$\overline{n}=(4\pi^{3/2})^{-1}$ for the Hooke's series. The idea
behind Eq.~(\ref{eq_vefflambda}) is that the average 
density $\overline{n}$
(and thus the average $\overline{s}_s$) is kept
fixed to mimic the fact that the one-electron density does not change along the
adiabatic connection while we turn on the e-e interaction.

\subsection{Linear adiabatic connection}
If we set $w^{\lambda}(s_{12})=\lambda/ s_{12}$, 
Eq.~(\ref{eq_vefflambda}) simply gives the Overhauser
potential of Eq.~(\ref{eq_ovscaled}) with a multiplying
factor $\lambda$ in front. From Eq.~(\ref{eq_y}), we see that this
corresponds to $E_c^{(2)}=\langle V_{ee}\rangle_c^{(2)}/2$,
as in the exact case. I.e., the simple approximation
of Eq.~(\ref{eq_vefflambda}) has the correct scaling behavior 
in the $\alpha\to 0$ limit. Our result for $E_c^{(2)}$ with the 
linear adiabatic connection thus gives an error of 4.5~mH for the
He series and 2~mH for the Hooke's series.

\subsection{The ``erf'' adiabatic connection}
A choice for  $w^{\lambda}$ that separates short- and long-range
effects is the ``erf'' adiabatic 
connection~\cite{adiabatic,erf,sav_madrid,julien,GS1},   
$w^{\lambda}(s_{12})=\erf(\lambda\,s_{12})/ s_{12}$,
for which Eq.~(\ref{eq_adiabatic}) becomes
\beq
E_c^{(2)}=\int_0^\infty d\lambda \int_0^\infty ds_{12}\, 
4\pi\,s_{12}^2\,\ft_c^{\lambda\;(1)}(s_{12})\,\frac{2}
{\sqrt{\pi}}\, e^{-\lambda^2\,s_{12}^2}.
\label{eq_erfconn}
\eeq 
The Overhauser-type potential corresponding to this interaction
is reported in the appendix of Ref.~\citen{GS1}. 
For the He isoelectronic series with $2\le Z\le 10$, the 
Overhauser-type approximation combined with the ``erf'' adiabatic
connection gives\cite{GS1} correlation energies with errors within 4~mH, 
better than the linear adiabatic connection that gives errors within 10~mH.

In the weakly-correlated limit, instead, we obtained, 
via Eq.~(\ref{eq_erfconn}),
$E_c^{(2)}=-0.041$~Ha for the He series and 
$E_c^{(2)}=-0.046$~Ha for the Hooke's series. The errors
with respect to the exact values, 6~mH and 4~mH, respectively, 
are thus slightly worse than those
obtained with the linear adiabatic connection.

\begin{figure}
\begin{center}
\includegraphics[width=7.cm]{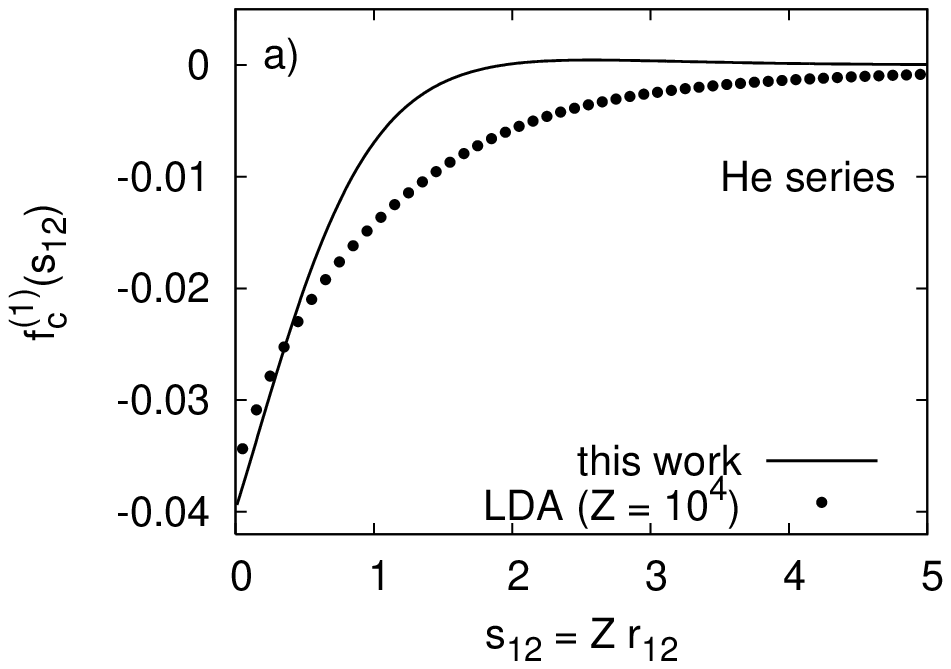}  

\includegraphics[width=7.cm]{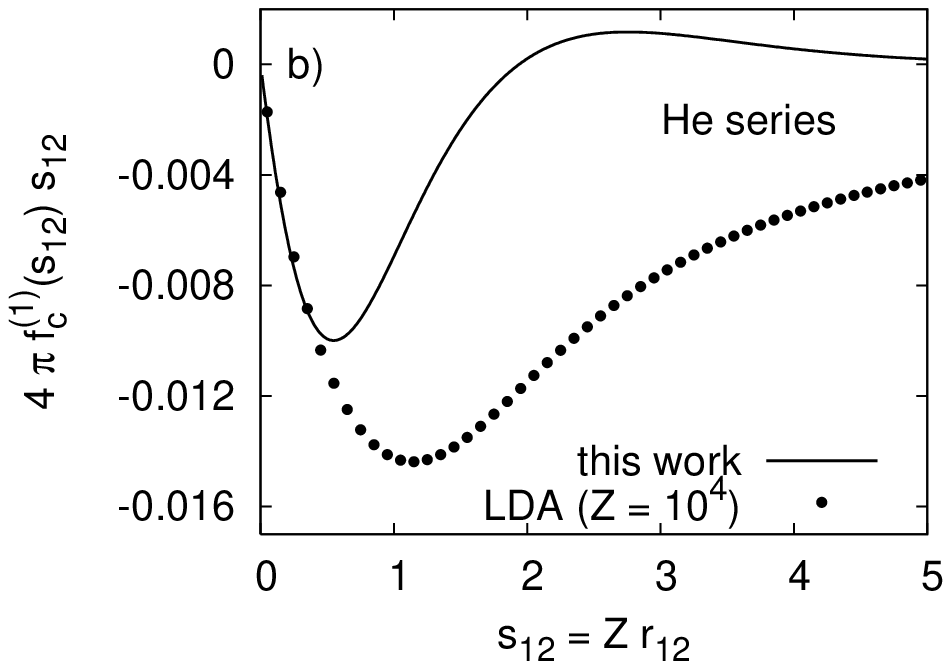}  
\end{center}
\caption{The correlated part of the intracule density, $\ft_c^{(1)}(s_{12})$, 
of order $\alpha=1/Z$ for the He series
[see Eq.~(\ref{eq_expf})]: the present calculation is compared with the LDA
approximation (panel a). Panel b shows the same quantities
multiplied by $4\pi s_{12}$: the integral under each curve gives
the correlation part of the second order contribution to the
expectation value  $\langle V_{ee}\rangle$, which diverges in the case of LDA.}
\label{fig_LDA_He}
\end{figure}

\begin{figure}
\begin{center}
\includegraphics[width=7.cm]{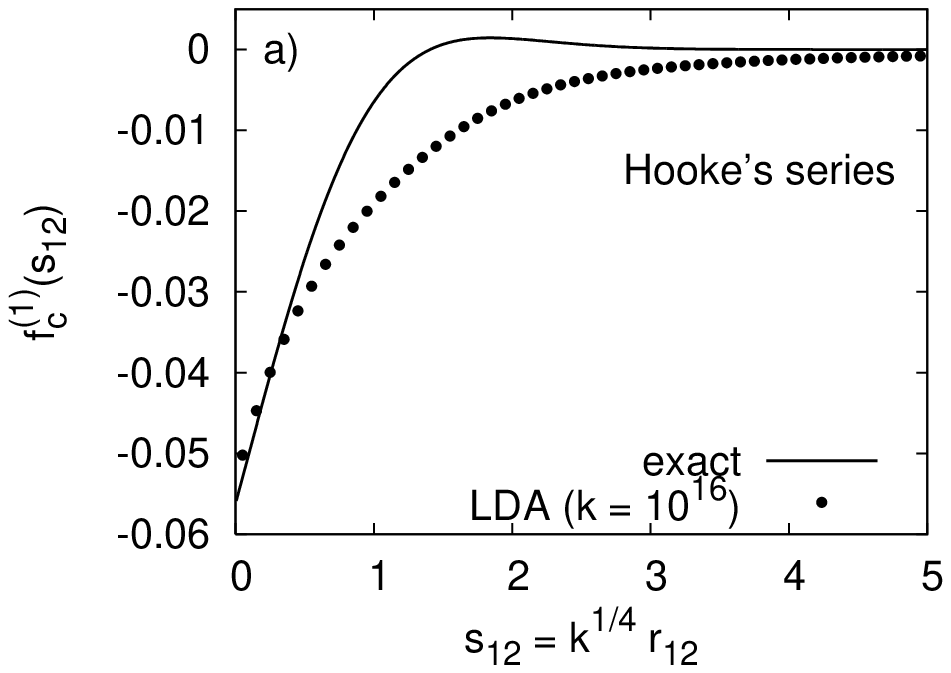}  

\includegraphics[width=7.cm]{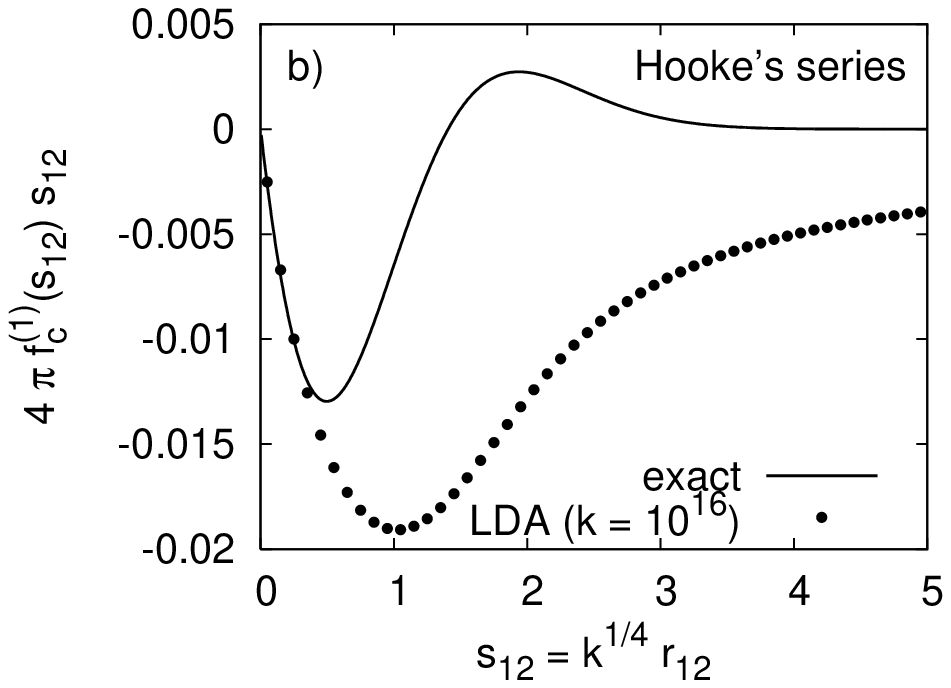}  
\end{center}
\caption{The correlated part of the intracule density, $\ft_c^{(1)}(s_{12})$,
 of order $\alpha=k^{-1/4}$ for the Hooke's atom series
[see Eq.~(\ref{eq_expf})]: the exact result is compared with the LDA
approximation (panel a). Panel b shows the same quantities
multiplied by $4\pi s_{12}$: the integral under each curve gives
the correlation part of the second order contribution to the 
expectation value  $\langle V_{ee}\rangle$, which diverges in the case of LDA.}
\label{fig_LDA_Hooke}
\end{figure}

\section{The LDA failure in the high-density limit:
an analysis from the intracule density}
\label{sec_LDA}
As a further element of comparison, we also computed the first-order 
$f^{(1)}_c(s_{12})$ within
the local-density approximation (LDA),
\beq
f^{(1){\rm LDA}}_c(s_{12})=\lim_{\alpha\to 0}
\frac{1}{\alpha}\int\frac{n^{(0)}(\sv)^2}{2}  
g_c\left(\tilde{k}_F(\sv)\, s_{12};\alpha\, \tilde{r}_s(\sv)\right) d\sv,
\label{eq_LDA}
\eeq 
where $g_c(r_{12};r_s)$ is the pair-correlation function of the 
uniform electron gas \cite{GP2} of density $n=(4\pi r_s^3/3)^{-1}$, and 
\beq
\tilde{k}_F(\sv) = [3\pi^2 n^{(0)}(\sv)]^{1/3}, \qquad
\tilde{r}_s(\sv)=\left[\frac{4\pi}{3}
n^{(0)}(\sv)\right]^{-1/3}.
\label{eq_rsforalpha}
\eeq
With these definitions, the density parameter $r_s$ of the uniform
electron gas is locally proportional to $\alpha$.
We have numerically evaluated the right-hand-side of Eq.~(\ref{eq_LDA})
at smaller and smaller $\alpha$ (i.e., at larger and larger
$Z$ and $k$), for $0\le s_{12}\le 5$. As $\alpha$ decreases, the
results tend to a well defined curve, shown in Figs.~\ref{fig_LDA_He}
and \ref{fig_LDA_Hooke}, 
together with the result from the Overhauser model (He series)
and the exact result (Hooke's series). 

Since, as shown by Eq.~(\ref{eq_LDA}), the $\alpha\to 0$ limit corresponds
to the $r_s\to 0$ limit of the uniform electron gas pair-correlation
function $g_c$,
to better understand the LDA result 
for $f_c$ we now analyze more in detail the
high-density behavior of $g_c$. This analysis extends and completes
the one done in Ref.~\citen{kieron}. When $r_s\to 0$,
the short-range part of $g_c$ scales as 
\beq
g_c(x,r_s\to 0)= r_s\, g_c^{(1)}(x)+O(r_s^2 \ln r_s), \qquad x=r_{12}/r_s, 
\label{eq_gcHD}
\eeq
where the function $g_c^{(1)}(x)$ does not depend explictily on
$r_s$ and has been computed by Rassolov 
{\it et al.}~\cite{rassolov}. It is accurately recovered by the
model $g_c$ of Ref.~\citen{GP2} that we have used in the
evaluation of Eq.~(\ref{eq_LDA}). The scaled variable $x$ is locally
proportional to the scaled variable $s_{12}$ [see Eq.~(\ref{eq_rsforalpha})].
Equation~(\ref{eq_gcHD}) thus shows
that the short-range part (corresponding to values of the scaled variable
$x$ not too large) of $g_c$ in the $r_s\to 0$ limit
has a scaling similar to the one
of the He and Hooke's series in the $\alpha\to 0$ limit. 
This is also reflected by a good performance of LDA for $s_{12}\lesssim 1$,
as shown by Figs.~\ref{fig_LDA_He} and \ref{fig_LDA_Hooke}.

However, the high-density electron gas is an extended system with
important long-range correlations that are not present in finite
systems like atoms and molecules. In fact,
the scaling of Eq.~(\ref{eq_gcHD}) is not valid when
$x\gg 1$: it has been shown that the long-range part of $g_c$ 
scales as~\cite{WP,PW92,GP2}
\beq
g_c(x\gg 1,r_s)\to r_s^2\, h(v),
\eeq 
where $v$ is another scaled variable, 
$v=\sqrt{r_s}\, x$, which is thus locally proportional to
$\sqrt{\alpha}\,s_{12}$. The function $h(v)$ has the following asymptotic
behaviors:
\beq
h(v\ll 1) \propto v^{-2}, \qquad h(v\gg 1) \propto v^{-4},
\eeq
which are also correctly included in the model $g_c$ of
Ref.~\citen{GP2}.
When $r_s\to 0$ (i.e., $\alpha\to 0$), even for very large
$x$ the scaled variable $v$ is small, so that the long-range
($x\gg 1$) behavior of $g_c$ is more and more dominated by the
small $v$ part of $h(v)$, i.e., it  behaves more and more
like $v^{-2}$ rather than like $v^{-4}$. It is this increasing
dominance  of the ``short-range component of the long-range part'' that causes
the $\propto\log(r_s)$ behavior in the correlation energy per electron of the
high-density electron gas, and thus the divergence of the LDA correlation
energy in the large-$Z$ and large-$k$ limit of the He and Hooke's 
atom sequences (see, e.g., Ref.~\citen{PMZ}). 
In fact, when $Z\to\infty$ (or $k\to\infty$), 
the high-density
long-range behavior of $g_c$ affects the long-range part of 
$f^{\rm LDA}_c(s_{12})$ in Eq.~(\ref{eq_LDA}).

The small-$v$ behavior $\propto v^{-2}$ of the function 
$h(v)$ is related to the
$1/r_{12}$ divergence of the Coulomb potential at {\em small} $r_{12}$.
For this reason, the $\propto\log(r_s)$ high-density behavior of the
correlation energy is still present in a uniform electron gas with
screened (or short-range only) Coulomb interaction
(e.g.,\cite{zecca} $\erfc(\lambda r_{12})/r_{12}$), while
is absent in an electron gas with long-range-only interaction
(e.g.,\cite{paziani,virialLR} $\erf(\lambda r_{12})/r_{12}$). 

\section{Conclusions}
\label{sec_conc}
We have computed the intracule density and the correlation energy
for the high-density (weakly-correlated)  limit of 
the He and Hooke's atom isoelectronic series via an 
approach\cite{GS1,GS2,GS4,nagy}
based on an ``average pair density functional theory'' (APDFT), and
inspired by the seminal work of Overhauser.\cite{Ov,GP1,pisani,CGPenagy2}
Unlike the currently available correlation energy 
functionals analyzed in Ref.~\citen{katriel}, the APDFT approach gives
accurate results for both series. In its present formulation,
the APDFT approach works well for two-electron systems and for
the uniform electron gas. Its extension to many-electron systems of
nonuniform density is a big challenge, and we are presently exploring
several different paths to achieve this ambitious goal.\cite{GS4}

We have also analyzed the LDA failure in the same 
weakly-correlated limit of the He and Hooke's atom series,  in terms
of the long-range part of the intracule density. 
The results of Katriel {\it et al.}\cite{katriel}
show that higher-order functionals such as PBE\cite{PBE} and
TPSS\cite{TPSS} can reasonably fix the LDA problems in the case of
the He isoelectronic series, but are much less satisfactory
for the Hooke's atom sequence, yielding a wrong scaling in the
$k\to\infty$ limit (PBE)  or a correct scaling with an
error of about 40\% on the asymptotic value of the correlation
energy (TPSS). As stressed by Katriel {\it et al.}\cite{katriel},
these differences in performances for the two series
raise serious doubts on the universality of currently available
correlation energy functionals. The accuracy of the results
obtained via the APDFT approach for both series suggests that
the effort towards its generalization  to many-electron
systems of nonuniform density could be really worthwhile.

\section*{Acknowledgments}
It is a pleasure to dedicate this methodological paper 
to Dennis Salahub, who did
pioneering work not only in the applications of DFT, but also in 
the exploration of new methodologies in the DFT framework.

\appendix
\section{$f_{\rm KS}^{(1)}(r_{12})$ for the He
isoelectronic series}
\label{app_fKS1}
For the He isoelectronic series
Eq.~(\ref{eq_fKS1}) corresponds to
\begin{eqnarray}
f_{\rm KS}^{(1)}(x) & = & \frac{1}{864\pi\,x}
\Big\{4 e^{-4x}[-41+3\,x\, (1+9\,x)]+ 81 e^{2x}(x-1)[{\rm Ei}(-6\,x)
 \nonumber \\ 
& &  -{\rm Ei}(-4\,x)]
 +e^{-2x}[164+27 (3+x\,(9+4\,x\,(3+2\,x)))
[{\rm Ei}(-2\,x) \nonumber \\
& &  -\gamma-\log(x)]
+ 3\,x\,[-163 
+ 6\,x\,(15+x\,(7+10\,x))-27\log(4/3)] \nonumber \\
& &  -162\log(2)+81\log(3)]\Big\},
\end{eqnarray}
where $\gamma$ and the function ${\rm Ei}$ have been defined
after Eq.~(\ref{eq_chi}).

\end{document}